\documentclass[9pt,twocolumn,twoside]{opticajnl}
\journal{opticajournal} % use for journal or Optica Open submissions
\setboolean{shortarticle}{false}
\usepackage{lineno}
\usepackage{graphicx}% Include figure files
\usepackage{dcolumn}% Align table columns on decimal point
\usepackage{amsmath,amsfonts,amssymb}
\usepackage{bm}% bold math
\usepackage{lipsum}
\usepackage[colorlinks=true, allcolors=blue]{hyperref}
\usepackage{url}
\usepackage{lineno}
\usepackage{booktabs}
\usepackage{tabularx}
\usepackage{amsmath}
\title{Quantum Key Distribution with a Negatively Charged Quantum Dot Single-Photon Source}

\author[1,*]{Parvendra Kumar}

\affil[1]{Optics  and  Photonics Centre, Indian Institute of Technology Delhi, Hauz Khas, New Delhi-110016, India}

\affil[*]{parvendra@iitd.ac.in}
\setboolean{displaycopyright}{false}

\begin{abstract}
Various quantum key distribution (QKD) protocols require bright single-photon sources with a very low probability of multiphoton emission. In this work, we investigate single-photon generation from a negatively charged quantum dot embedded in an elliptical pillar microcavity, driven by either resonant excitation or adiabatic rapid passage (ARP). Our results show that ARP excitation significantly suppresses multiphoton emission and improves photon indistinguishability compared to resonant excitation. We further evaluate the secure key rates of both BB84 and twin-field QKD (TF-QKD) using a quantum-dot single-photon source and compare its performance with that of Poisson-distributed sources (PDS), such as weak coherent pulses. The analysis reveals that ARP provides a modest but consistent enhancement in secure key rate relative to resonant excitation. We show that quantum-dot single-photon sources offer improved performance compared to PDS sources at short and intermediate distances within the chosen parameter regime; however, at longer distances, PDS sources slightly surpass them in both decoy-state BB84 and TF-QKD under the same conditions.
\end{abstract}

\begin{document}

\maketitle

\section{Introduction}\label{sec:level1}
The distribution of secret cryptographic keys between the distant parties using the fundamental principles of quantum mechanics is a crucial step for exchanging the information with unconditional security. This process is generally known as quantum key distribution (QKD). Since the first proposal by Charles Bennett and Gilles Brassard in 1984 (BB84), QKD has evolved from a theoretical concept to a full-fledged research area with practical applications ~\cite{ref1,ref2,ref3, ref4, ref5, ref6}. The secret keys can be established by appropriately distributing and measuring the quantum bits between two legitimate parties ~\cite{ref7, ref8, ref9,ref10}. Quantum bits (qubits) can be realized in the different degrees of freedom of single-photon or entangled-photon states. These quantum states are produced via various platforms, including atoms, nonlinear crystals, and semiconductor quantum dots, and can be used to construct QKD methods ~\cite{ref11,ref12,ref13,ref14,ref15}. The quantum principles, like the no-cloning theorem and the disturbance caused by quantum measurements, provide the information-theoretic security since any eavesdropping attempt is reflected as an increase in the quantum bit error rate (QBER). Accordingly, the key distribution protocols can be aborted if the QBER increases beyond a threshold value. Generally, most of the experimental works of QKD employ the weak coherent sources. However, such sources are inherently Poissonian, which means that the weak pulses occasionally contain multiple photons. This situation opens the window for photon-number-splitting (PNS) attacks, which severely restrict both secure communication distance and key rate ~\cite{ref16, ref17, ref18, ref19, ref20}. However, this problem can be overcome by adding multiple intensity levels—specifically, signal and decoy states—which allow the infinite and realistic decoy-state BB84 protocols and enable Alice and Bob to precisely estimate the single-photon yield and error rate, thereby restoring security that is comparable to that of an ideal single-photon source ~\cite{ref21,ref22,ref23,ref24,ref25}. Towards this end, there has been a surge in experimental demonstrations of QKD using quantum dot-based single-photon sources, owing to improved source and conversion efficiencies ~\cite{ref26,ref27,ref28, ref29, ref30}. Moreover, recently QKD has also been demonstrated with imperfect single-photon sources ~\cite{ref31, ref32}.   

Another significant development is measurement-device-independent quantum key distribution (MDIQKD) alongside twin-field quantum key distribution (TF-QKD) ~\cite{ref33, ref34}. Notably, TF-QKD addresses the fundamental linear rate-distance constraint that impacts point-to-point QKD ~\cite{ref34}. Specifically, TF-QKD allows for much greater operating distances by achieving a key rate that scales with the square root of channel transmittance, rather than linearly, through the interference of weak optical beams from two distant users at an untrusted central station. But TF-QKD's weak coherent states are vulnerable to PNS assaults once more. These vulnerabilities are eliminated by integrating the decoy-state approach with TF-QKD, which offers precise estimation of single-photon contributions. Without using quantum repeaters, secure key distribution over fiber lines longer than 800 km has been made possible via decoy-state TF-QKD and its variations ~\cite{ref35,ref36,ref37,ref38}.

In this work, we theoretically investigate the single-photon generation from a negatively charged quantum dot driven by resonant excitation and adiabatic rapid passage (ARP) to investigate the performance of quantum key distribution (QKD). Our results show that ARP excitation enables higher source brightness while significantly suppressing multiphoton emission compared to conventional resonant driving. Using these improved source characteristics, we evaluate the secure key rate (SKR) achievable in the BB84 protocol without and with decoy states, as well as in the original twin-field QKD (TF-QKD) protocol with decoy states. We observe that the SKR value achieved with the quantum-dot single-photon source exceeds the values obtained with PDS at short and intermediate transmission distances for both BB84 and TF-QKD. Furthermore, the enhanced brightness and reduced multiphoton contribution provided by ARP lead to a consistent additional improvement in the SKR compared to resonant excitation. While previous studies have investigated quantum cryptography using neutral quantum-dot emitters under various excitation schemes ~\cite{ref39}, the present work focuses instead on an adiabatically driven, negatively charged quantum dot embedded in an elliptical micropillar cavity and operating at telecommunication wavelengths ~\cite{ref40,ref41}. We show that the ellipticity-induced nondegeneracy of orthogonally polarized modes in elliptical pillar-microcavity enables the  enhanced emission of the quantum dot into the resonant cavity mode, thereby improving the brightness together with reduced two-photon emission probability. It is worthwhile to mention that in TF-QKD, the bit value is encoded through the phase relationship between the vacuum state and the single-photon state. Consequently, the implementation of TF-QKD would be unfeasible even with ideal single-photon sources exhibiting a unit probability of single-photon emission, as there exists no relative phase to encode the bit information ~\cite{ref34}. However, quantum dot sources enable the generation of the necessary superposition of vacuum and single-photon states by adjusting either the Rabi frequency or the temporal duration of the phase-modulated pumping pulse ~\cite{ref42,ref43,ref44}.

\section{Theory: Modelling of single-photon generation}{\label{sec2}}

\begin{figure}[htbp]
\centering
\includegraphics[width=\linewidth]{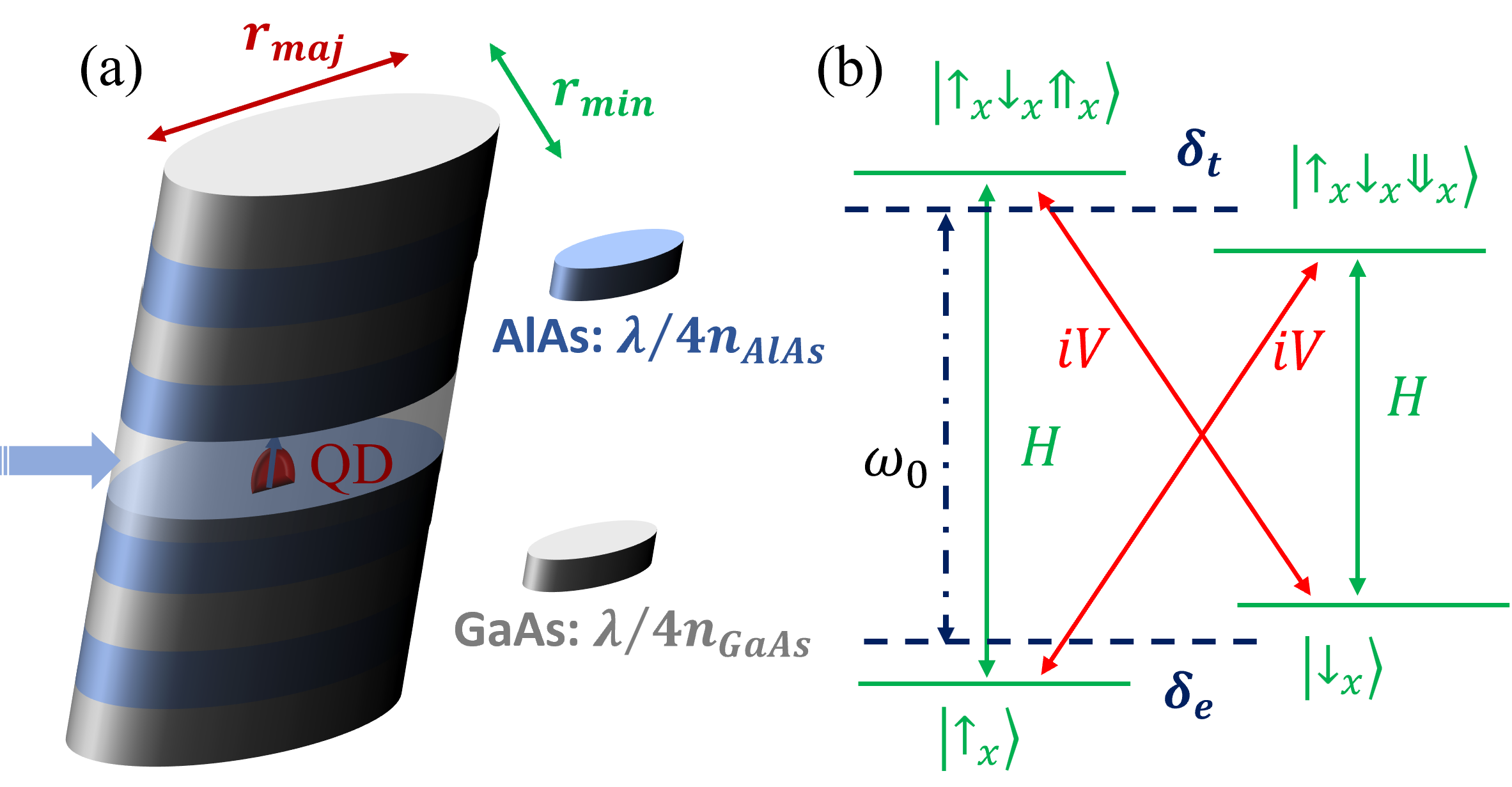}
\caption{(a) Illustration of an elliptical pillar microcavity comprising GaAs/AlAs distributed Bragg reflectors with an embedded negatively charged quantum dot (QD). The orientation of the cavity is intentionally tilted to improve the visualization of the elliptical cross-section. (b) Energy-level diagram and optical transitions of a four-level QD system induced by an externally supplied magnetic field in the Voigt configuration. A description of the different energy levels and optical transitions is provided in the text. }
\label{fig:1}
\end{figure}

We consider in-plane excitation of a negatively charged self-assembled quantum dot (QD) in an elliptical pillar microcavity ~\cite{ref28}. Alternative platforms, such as photonic crystals and circular Bragg grating bullseye cavities, can also be utilized. However, the elliptical pillar microcavity can provide a favorable balance between collection efficiency and indistinguishability ~\cite{ref28, ref45, ref46}. The number of charge carriers in self-assembled quantum dots (QDs) is typically introduced by embedding the QD within a voltage-tunable diode or field-effect structure. This setup allows the applied bias to shift the energy levels of the dot relative to those of a carrier reservoir, a two-dimensional electron gas ~\cite{ref47}.
In the microcavity, the number of distributed Bragg reflectors (DBRs) on the bottom side is assumed to exceed that on the top side, thereby enabling preferential photon emission from the top of the microcavity~\cite{ref48}.
The spin states of a single confined electron along the $x$-axis are represented by
$|1\rangle = |\uparrow_x\rangle = \frac{1}{\sqrt{2}}\left(|\uparrow_z\rangle + |\downarrow_z\rangle\right)$
and
$|2\rangle = |\downarrow_x\rangle = \frac{1}{\sqrt{2}}\left(|\uparrow_z\rangle - |\downarrow_z\rangle\right)$.
Similarly, the trion states are given by
$|3\rangle = |\uparrow_x \downarrow_x \Downarrow_x\rangle$
and
$|4\rangle = |\uparrow_x \downarrow_x \Uparrow_x\rangle$,
where $\Uparrow_x$ and $\Downarrow_x$ represent the heavy-hole spin states along the positive and negative $x$-axis, respectively.
The two ground states $|1\rangle$ and $|2\rangle$ are optically coupled to the excited states $|3\rangle$ and $|4\rangle$ via horizontally (H) and vertically (V) polarized laser fields, as depicted in Fig. ~\ref{fig:1} (b).
Here, $\omega_0$ denotes the energy separation between the ground and excited states in the absence of an externally applied magnetic field. The spin and trion states are energetically split due to the Zeeman energies
$\delta_e = g_e \mu_B B$ and $\delta_t = g_h \mu_B B$, respectively.
Here, $g_e$ and $g_h$ are the Land\'e $g$-factors of the electron and heavy hole, $\mu_B$ is the Bohr magneton, and $B$ is the externally applied magnetic field along the $x$-axis.

The dynamics of the spin states coupled to orthogonally polarized cavity modes and driven by a horizontally polarized laser pulse are described by the Hamiltonian $H = H_0 + H_{\mathrm{int}}$, where $H_0$ represents the free Hamiltonian of the quantum dot and cavity modes, and $H_{\mathrm{int}}$ describes their interaction.
These are given by~\cite{ref49} as
$H_0 = -\frac{\delta_e}{2}\sigma_{11} + \frac{\delta_e}{2}\sigma_{22} + \left(\omega_0 - \frac{\delta_t}{2}\right)\sigma_{33} + \left(\omega_0 + \frac{\delta_t}{2}\right)\sigma_{44} + \omega_h a^\dagger a + \omega_v b^\dagger b$,
and
$H_{\mathrm{int}} = g_h a^\dagger(\sigma_{14} + \sigma_{23}) + i g_v b^\dagger(\sigma_{24} + \sigma_{31}) + \Omega_h e^{i(\omega_l t + \alpha t^2)}(\sigma_{23} + \sigma_{14}) + \mathrm{H.c.}$ Here, $\omega_h$ and $\omega_v$ are the frequencies of the H- and V-polarized cavity modes with annihilation operators $a$ and $b$, respectively, and $\sigma_{ij} = |i\rangle\langle j|$.
The parameters $g_h$ and $g_v$ denote the coupling strengths between the cavity modes and the quantum dot transitions.

The Rabi frequency describing the coupling between the laser pulse and the H-polarized transitions is given by $\Omega_h(t) = \mu E(t)/\hbar$, where $\mu$ is the electric dipole moment and the electric field envelope is
$E(t) = E_0 \exp\!\left[-(t - 2.5\tau_p)^2/\tau_p^2\right]$.
The full width at half maximum of the pulse is $\tau_{\mathrm{FWHM}} = 1.177\,\tau_p$, and $\alpha$ is the linear chirp parameter.

Transforming into a frame rotating at the laser frequency $\omega_l$, the Hamiltonian becomes
$H_r = U^\dagger H U + i(\partial U^\dagger/\partial t)U$, with
$U = \exp\!\left[-i\omega_l t (a^\dagger a + b^\dagger b + \sigma_{33} + \sigma_{44})\right]$.
This yields $H_r = H_0^r + H_{\mathrm{int}}^r$, where
$H_0^r = -\frac{\delta_e}{2}\sigma_{11} + \frac{\delta_e}{2}\sigma_{22} + \left(\Delta - \frac{\delta_t}{2}\right)\sigma_{33} + \left(\Delta + \frac{\delta_t}{2}\right)\sigma_{44} + \Delta_{hl} a^\dagger a + \Delta_{vl} b^\dagger b$
and
$H_{\mathrm{int}}^r = g_h a^\dagger(\sigma_{14} + \sigma_{23}) + i g_v b^\dagger(\sigma_{24} + \sigma_{31}) + \Omega_h e^{i\alpha t^2}(\sigma_{23} + \sigma_{14}) + \mathrm{H.c.}$ Here, $\Delta = \omega_0 - \omega_l$, $\Delta_{hl} = \omega_h - \omega_l$, $\Delta_{vl} = \omega_v - \omega_l$, and $\Delta_{HV} = \omega_h - \omega_v$. In this work, we assume that the quantum dot is pre-initialized in its spin state $|\downarrow_x\rangle$. This state can be prepared via optical pumping using few-nanosecond or sub-nanosecond laser pulses, followed by rapid relaxation of the trion state. Notably, the long lifetime of electron spin states, ranging from milliseconds to seconds, enables optical initialization of spin states with a near-unity-fidelity  ~\cite{ref50, ref51, ref52}.

To investigate single-photon generation, we numerically solve the master equation
$d\rho/dt = -i[H_r,\rho] + \mathcal{L}_{\mathrm{cav}}\rho + \mathcal{L}_{\mathrm{QD}}^r\rho + \mathcal{L}_{\mathrm{QD}}^d\rho$,
where
$\mathcal{L}_{\mathrm{cav}}\rho = (\kappa/2)\left[\mathcal{L}(a)\rho + \mathcal{L}(b)\rho\right]$,
$\mathcal{L}_{\mathrm{QD}}^r\rho = (\Gamma_{14}/2)\mathcal{L}(\sigma_{14})\rho + (\Gamma_{24}/2)\mathcal{L}(\sigma_{24})\rho + (\Gamma_{13}/2)\mathcal{L}(\sigma_{13})\rho + (\Gamma_{23}/2)\mathcal{L}(\sigma_{23})\rho$,
and
$\mathcal{L}_{\mathrm{QD}}^d\rho = (\gamma_{33}/2)\mathcal{L}(\sigma_{33})\rho + (\gamma_{44}/2)\mathcal{L}(\sigma_{44})\rho$,
with the Lindblad superoperator defined as
$\mathcal{L}(o)\rho = 2o\rho o^\dagger - o^\dagger o\rho - \rho o^\dagger o$.

\section{Results and discussions}{\label{sec3}}
We investigate the single-photon parameters, including indistinguishability, brightness, and the multiphoton emission probability, by numerically solving the master equation together with the quantum regression theorem. In the simulations, we choose typical parameters for the quantum dot and cavity system: $\Gamma_{14} = \Gamma_{24} = \Gamma_{13} = \Gamma_{23} = 1~\mu\mathrm{eV}$, $\gamma_{33} = \gamma_{44} = 2~\mu\mathrm{eV}$, $g_e = 0.378$, $g_h = 0.202$, $\mu_B = 0.578~\mu\mathrm{eV}$, and $B = 5~\mathrm{T}$~\cite{ref49}. The remaining parameters are chosen as $g_h = g_v = 60~\mu\mathrm{eV}$ for resonant excitation and $g_h = g_v = 70~\mu\mathrm{eV}$ for adiabatic excitation, $\alpha = 7~\mathrm{ps}^{-2}$ for adiabatic excitation and $\alpha = 0$ for resonant excitation, and a cavity decay rate of $\kappa = 150~\mu\mathrm{eV}$.
\subsection{Indistinguishability and Brightness of Single-Photon}
The purity of the emitted photon’s quantum state is quantified by its single-photon indistinguishability. When the probability of emitting more than one photon is negligible, indistinguishability reflects the first-order (field) coherence of the source, indicating that the spectra of the emitted photon wave packets are mutually identical. However, when the source exhibits a nonzero probability of multiphoton emission, the indistinguishability is no longer determined solely by first-order coherence. In such cases, particularly for pulse-triggered single-photon sources, it also depends critically on the second-order (intensity) coherence, since multiphoton contributions degrade the overall purity of the quantum state. Experimentally, a Mach-Zehnder interferometer can be used to measure the indistinguishability of a single-photon source, triggered with a delay much longer than the lifetime of the excited states  ~\cite{ref53}. 
\begin{figure}[!t]
\centering
\includegraphics[width=\linewidth]{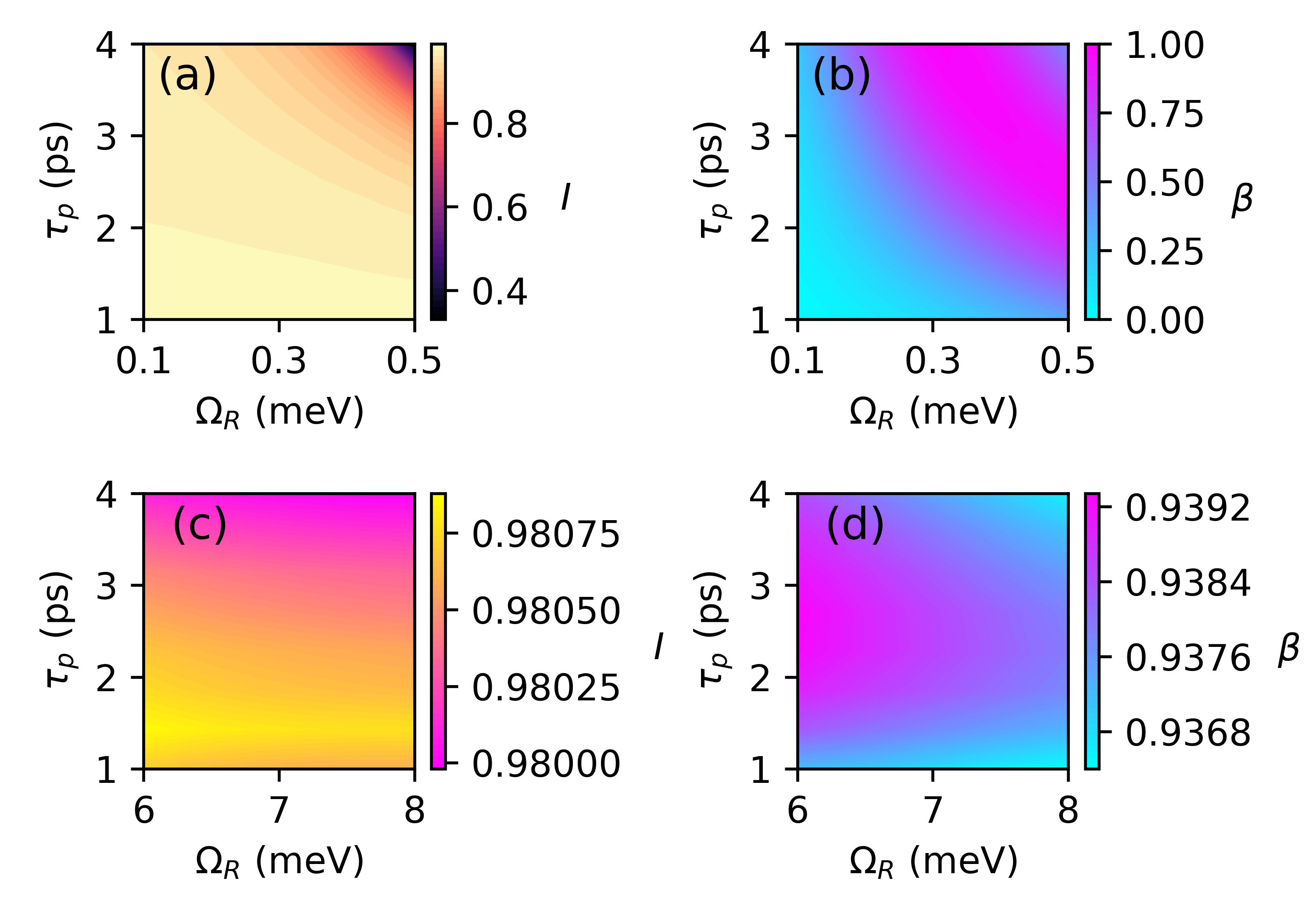}
\caption{Evolution of indistinguishability $I$ and brightness $\beta$ as functions of Rabi frequency and pulse duration. (a,b) Resonant excitation. (c,d) Adiabatic excitation. $\Delta_{HV} = 0.6~\mathrm{meV}$.}
\label{fig:2}
\end{figure}

In Fig.~\ref{fig:2}, we present the single-photon indistinguishability $I$ and brightness $\beta$ as functions of the Rabi frequency and pulse duration for resonant excitation ($\alpha = 0$) and adiabatic excitation ($\alpha = 7~\mathrm{ps}^{-2}$). The mathematical expressions used to compute these quantities are provided in Appendix~\ref{A}. From Figs.~\ref{fig:2} (a) and ~\ref{fig:2} (c), it is evident that adiabatic excitation yields a significantly higher indistinguishability, exceeding 0.98, compared to the value of approximately 0.92 obtained under resonant excitation. This improvement primarily arises from the substantially reduced multiphoton emission probability for adiabatic excitation, as discussed in the next section. Moreover, as expected, the indistinguishability achieved through adiabatic rapid passage is considerably more robust against variations in both the Rabi frequency and the pulse duration, whereas the resonant scheme shows a stronger sensitivity to these parameters. Similarly, the brightness obtained under adiabatic excitation is not only higher but also more robust and stable compared to that achieved with resonant excitation, as evident from Figs. ~\ref{fig:2} (b) and ~\ref{fig:2} (d).

\begin{figure}[htbp]
\centering
\includegraphics[width=0.85\linewidth]{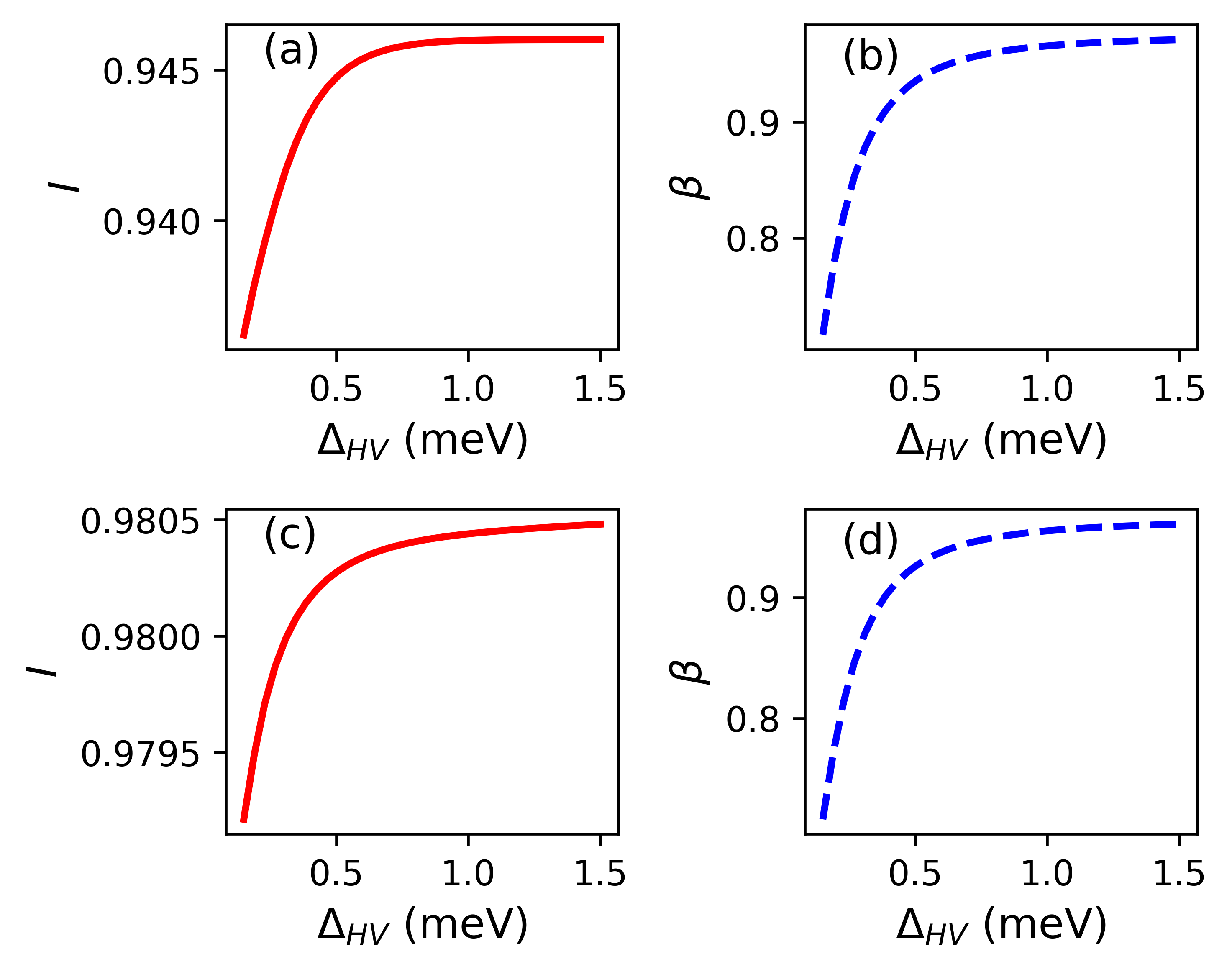}
\caption{Evolution of the indistinguishability $I$ and brightness $\beta$ as a function of the detuning between orthogonally polarized cavity modes. Panels (a,b) show resonant excitation, while (c,d) correspond to adiabatic excitation.}
\label{fig:3}
\end{figure}

Next, in Fig. ~\ref{fig:3}, we investigate the indistinguishability $I$ and brightness $\beta$ as functions of $\Delta_{HV}$ for resonant and adiabatic excitation. In both cases, $I$ and $\beta$ increase with increasing $\Delta_{HV}$. This behavior arises because, for larger $\Delta_{HV}$, the emission from state $|3\rangle$ is preferentially funneled into the V-polarized cavity mode, which is resonant with the $|1\rangle \leftrightarrow |3\rangle$ transition. At the same time, the probability of emission into the H-polarized cavity mode, mediated by the $|2\rangle \leftrightarrow |3\rangle$ and $|1\rangle \leftrightarrow |4\rangle$ transitions, is significantly suppressed. As a result, multiphoton emission events are reduced, leading to enhanced single-photon indistinguishability and brightness. The detuning $\Delta_{HV}$ between orthogonally polarized cavity modes can be adjusted by modifying the ellipticity of the micropillar cavity ~\cite{ref48}.

\subsection{Probability of Multiphoton Emission and Secure Key Rate}
Next, we analyze the probabilities of single- and two-photon emission from the quantum-dot source under both resonant and adiabatic excitation, and compare the resulting secure key rate (SKR) with those obtained using quantum dot source (QDS) and PDS sources. Throughout this work, we have neglected the probability of generating three or more photons, as the $n$-photon generation probability for three-level systems scales as $(\Gamma~\tau_{\mathrm{FWHM}})^{2(n-1)}$~\cite{ref54}. For the chosen parameters, the probability of three-photon generation, $p_3$, scales as $10^{-9}$, which is negligibly small. The detailed expressions for the multiphoton emission probabilities and the SKR for BB84 and TF-QKD are summarized in Appendix~\ref{A} and~\ref{B}, respectively. From Fig.~\ref{fig:4}, it is evident that the single-photon generation probability increases, while the two-photon emission probability simultaneously decreases with increased $\Delta_{\mathrm{HV}}$ for both resonant and adiabatic excitation.

\begin{figure}[htbp]
\centering
\includegraphics[width=0.8\linewidth]{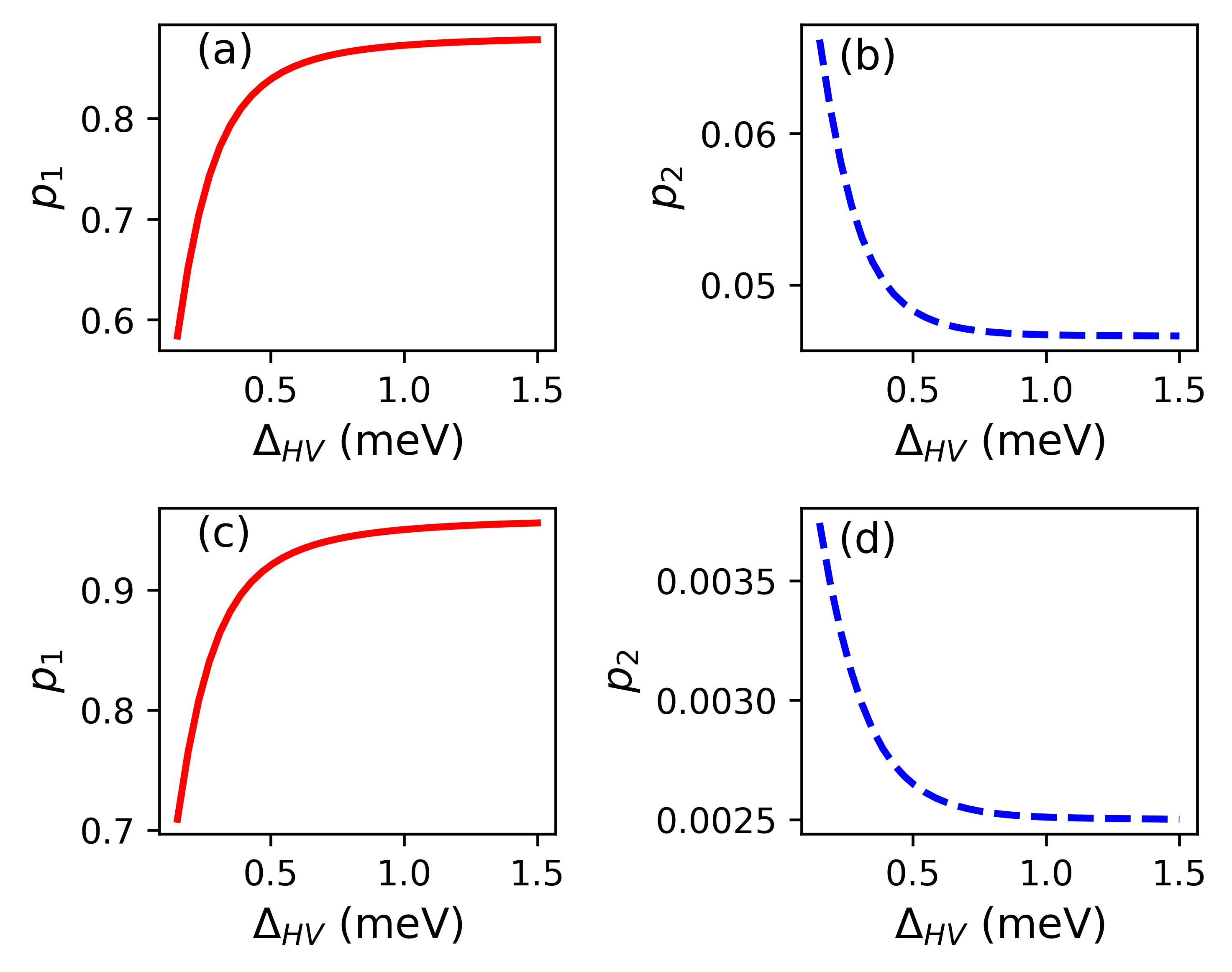}
\caption{Evolution of one-photon $p_{1}$ and two-photon probabilities $p_{2}$ as a function of the detuning between orthogonally polarized cavity modes. Panels (a,b) show resonant excitation, while (c,d) correspond to adiabatic excitation.}
\label{fig:4}
\end{figure}

\begin{figure}[!t]
\centering
\includegraphics[width=0.8\linewidth]{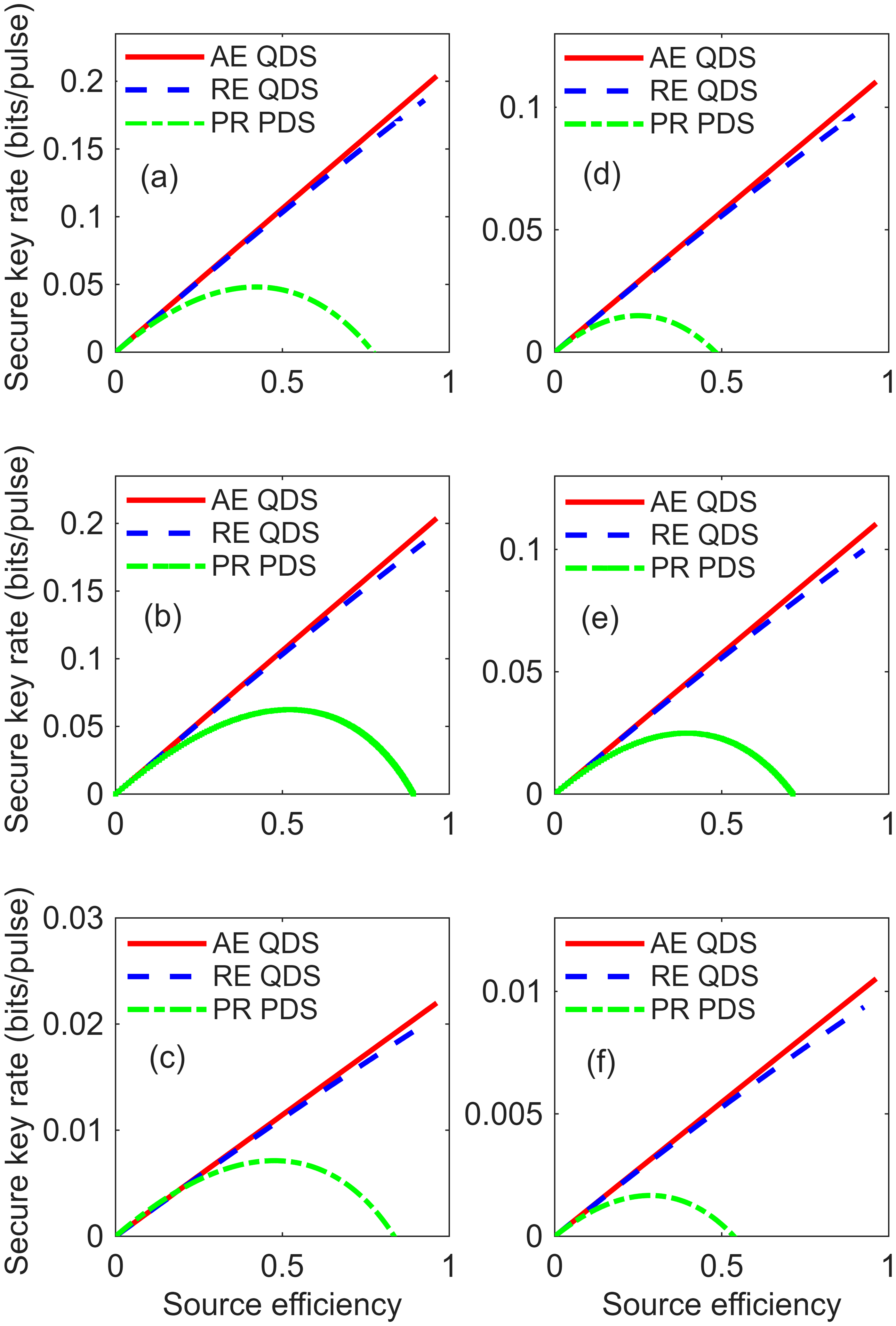}
\caption{Evolution of the secure key rate as a function of source efficiency for three QKD schemes using adiabatic-excitation quantum-dot single-photon sources (AE-QDS, red solid lines), resonant-excitation quantum-dot single-photon sources (RE-QDS, blue dashed lines), and phase-randomized Poissonian sources (PR-PDS, green dash-dotted lines). 
Panels (a,d) correspond to BB84 without decoy states, (b,e) to BB84 with decoy states, and (c,f) to twin-field QKD with decoy states, for a transmission distance of $L=10~\mathrm{km}$. 
The source efficiency is defined as $1-e^{-\mu}$ for PDS and $1-\sum_{n=0}^{\infty} p_n (1-\eta)^n$ for QDS, where $p_n$ is the $n$-photon emission probability and $\eta=50\%$ is the collection efficiency. 
Simulation parameters for panels (a--c) are $e_d=2\%$, $Y_0=10^{-9}$, and $\eta_d=100\%$, while for panels (d--f) we chose experimentally accessible parameters, $e_d=4\%$, $Y_0=10^{-5}$, and $\eta_d=80\%$.
}
\label{fig:5}
\end{figure}

\begin{figure*}[!ht]
    \centering
    \includegraphics[width=.9\linewidth]{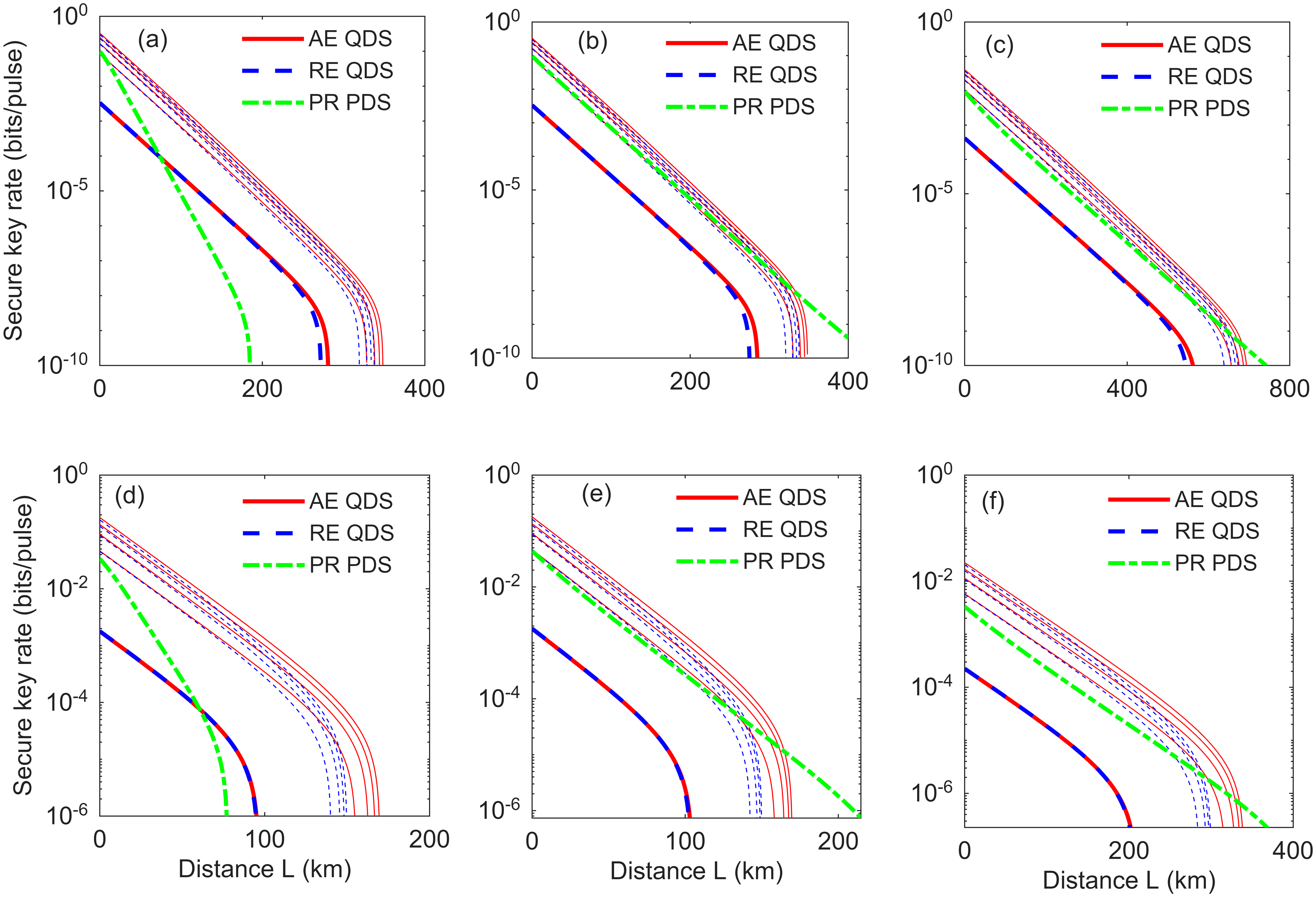}
    \caption{
Evolution of the secure key rate as a function of distance for three QKD schemes using adiabatic-excitation quantum-dot single-photon sources (AE-QDS, red solid lines), resonant-excitation quantum-dot single-photon sources (RE-QDS, blue dashed lines), and phase-randomized Poissonian sources (PR-PDS, green dash-dotted lines). 
Panels (a,d) correspond to BB84 without decoy states, (b,e) to BB84 with decoy states, and (c,f) to twin-field QKD with decoy states. 
Simulation parameters for panels (a--c) are $e_d=2\%$, $Y_0=10^{-9}$, and $\eta_d=100\%$, while for panels (d--f) they are $e_d=4\%$, $Y_0=10^{-5}$, and $\eta_d=80\%$. 
The optimal mean photon numbers are chosen as $\mu=0.7\,\eta_d \eta_t$ for panels (a,b), $\mu=0.5$ for panels (c,d), $\mu=0.765$ for panel (e), and $\mu=0.365$ for panel (f).
}
\label{fig:6}

\end{figure*}

Notably, the two-photon emission probability under adiabatic excitation is at least an order of magnitude lower than that achieved with resonant excitation. Consequently, the performance of the quantum-dot source under adiabatic excitation is slightly superior to that under resonant excitation, as illustrated in Figs.~\ref{fig:5} and~\ref{fig:6}.

To compare the performance of QDS and PDS sources for BB84 QKD (without decoy and with decoy states) and TF-QKD with decoy states, we plot the SKR per pulse as a function of the source efficiency for two different set of parameters.

\begin{table}[htbp]
\centering
\caption{Simulation parameters of BB84 and Twin Field QKD}
\renewcommand{\arraystretch}{1.0}

\begin{tabularx}{\columnwidth}{X c c}
\toprule
\textbf{Description} & \textbf{Parameter} & \textbf{Value} \\
\midrule
Alignment error rate & $ e_{d}$ & 2$\%$ \\
Basis reconciliation factor & $q$ & $0.5$ \\
Dark count probability & $Y_{0}$ & $10^{-9}$ \\
Error correction inefficiency & $f$ & $1.2$ \\
Zero-photon probability for RE and AE & $p_{0}$ & $7.4\%$ and $3.75\%$ \\
One-photon probability for RE and AE & $p_{1}$ & $88\%$ and $96\%$ \\
Two-photon probability for RE and AE & $p_{2}$ & $4.6\%$ and $0.25\%$ \\
Intrinsic error rate & $e_{s}$ & $2\%$ \\
Number of phase slices & $M$ & $16$ \\
Duty cycle & $d$ & $1$ \\
Fiber loss coefficient & $\alpha$ & $0.21~\mathrm{dB/km}$ \\
\bottomrule
\end{tabularx}

\end{table}

As shown in Fig.~\ref{fig:5}, the SKR for PDS source reaches a maximum at an optimal source efficiency; beyond this point, the increasing contribution of multiphoton pulses leads to a rapid degradation of the SKR, eventually driving it to zero. In contrast, for QDS under both adiabatic excitation (AE) and resonant excitation (RE), the SKR increases nearly linearly with source efficiency because the vacuum and single-photon components remain dominant across the entire efficiency range. Notably, the AE-driven QDS achieves a slightly higher SKR than the RE-driven QDS, owing to its reduced multiphoton emission probability. Moreover Figs. 5 (d)–(f) indicate that both the secure key rate and transmission distance decrease when using widely accessible parameters, including detection efficiency, dark counts and alignment error rate.

\begin{figure}[htbp]
\centering
\includegraphics[width=0.9\linewidth]{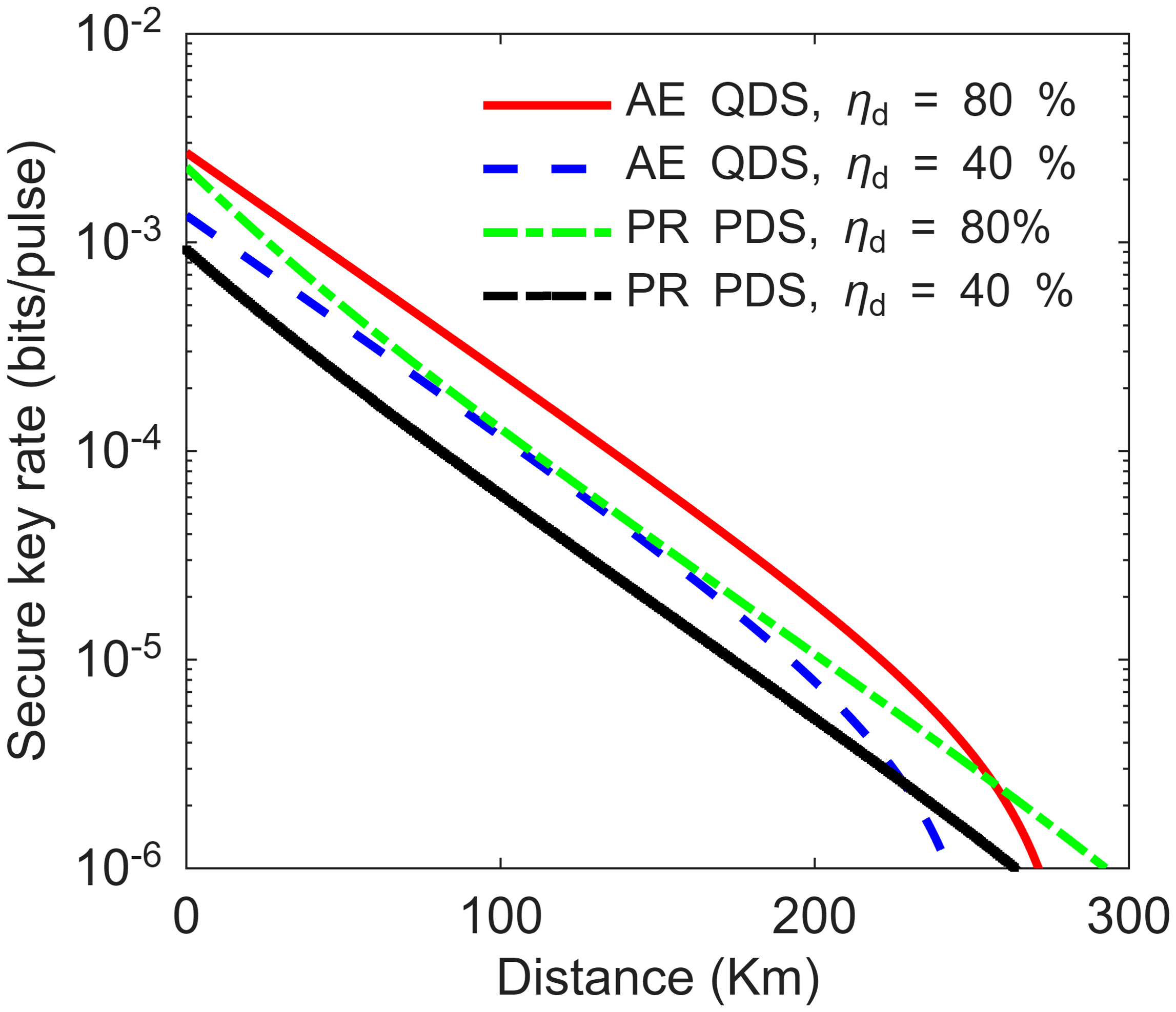}
\caption{Evolution of the TF-QKD secure key rate as a function of transmission distance for a quantum-dot source under adiabatic excitation with a collection efficiency of $20\%$, and for phase-randomized Poissonian coherent sources, at a detector efficiency of $\eta_d=80\%$ and dark count probability $Y_0=10^{-5}$. The optimal mean photon number is chosen as $\mu=0.365$. }
\label{fig:7}
\end{figure}

In Fig.~\ref{fig:6}, we present the performance of the QDS pumping schemes for collection efficiencies ranging from $1\%$ to 100$\%$ with intermediate curves corresponding to increments of $20\%$, and compare these results with the optimal performance of phase-randomized (PR) PDS at its best choice of $\mu$. 
As shown in Fig.~\ref{fig:6} (a) and (d), the SKR achieved with QDS is consistently higher, enabling secure communication over significantly longer distances. 
Even with a collection efficiency of only 1\%, QDS supports a substantially longer transmission distance than PR-PDS; this advantage increases to at least a threefold improvement when the collection efficiency reaches 100\%.
Furthermore, Figs.~\ref{fig:6}(b,e) and (c,f) show that the SKR obtained with QDS exceeds that of PDS over short and intermediate distances, even for a collection efficiency of $20\%$. Such efficiencies are achievable with state-of-the-art sources exhibiting $60$--$70\%$ collection efficiencies in the near-infrared (NIR) regime~\cite{ref26,ref27,ref28}, combined with a telecom-wavelength conversion efficiency of $\sim 35\%$~\cite{ref29}. However, at longer distances, PDS sources eventually marginally outperform quantum-dot sources in both decoy-state BB84 and TF-QKD. Moreover, Figs.~\ref{fig:6} (d--f) show that both the secure key rate and the maximum transmission distance decrease for more realistic parameters, namely $e_d = 4\%$, $Y_0 = 10^{-5}$, and $\eta_d = 80\%$. Under these conditions, the optimal mean photon number for TF-QKD with a PDS is found to be $\mu = 0.365$.
 
Finally, to assess the practical performance of TF-QKD, we investigate the dependence of the secure key rate on detector efficiency for both PDS and QDS sources. As shown in Fig.~\ref{fig:7}, the secure key rate increases with detector efficiency. Detection efficiency comparable to those of superconducting nanowire single-photon detectors (SNSPDs), typically exceeding $80\%$, enables higher key rates and longer transmission distances. Moreover, adiabatically driven QDS maintains superiority over PDS sources at short and intermediate distances for both chosen detection efficiencies.

\section{Conclusion}{\label{sec4}}
In conclusion, we numerically investigate a quantum-dot single-photon source based on a negatively charged emitter embedded in an elliptical pillar microcavity. We show that the spectral nondegeneracy of the cavity modes effectively suppresses two-photon emission for both adiabatic and resonant excitation, thereby enhancing secure key rates. Among the two schemes, adiabatic excitation provides improved robustness and brightness stability while reducing the multiphoton probability by an order of magnitude compared to resonant excitation. Within the chosen parameter regime, the generated quantum-dot single photons enable enhanced secure key rates for both BB84 and twin-field QKD in the infinite decoy-state limit at short and intermediate transmission distances. However, at longer distances, Poissonian sources exhibit a marginal advantage under the same conditions. We further analyze the dependence of the secure key rate on quantum channel distance under realistic conditions, including detection efficiency, dark count probability, and alignment errors. These results provide practical insight into the performance of negatively charged quantum-dot single-photon sources for BB84 and Twin-field QKD and can be leveraged in future quantum communication experiments.

\appendix

\section{ APPENDIX A: Indistinguishability and Brightness of Single-Photon}{\label{A}}
The indistinguishability of photons emitted into the vertically polarized mode is numerically calculated using~\cite{ref55} as $I = 1 - \frac{\int_0^T dt \int_0^T d\tau ,[G_{\rm pop}^2(t,\tau) + g^2(t,\tau) - |g^1(t,\tau)|^2]}{\int_0^T dt \int_0^T d\tau ,[2G_{\rm pop}^2(t,\tau) - |\langle b(t+\tau)\rangle \langle b^\dagger(t)\rangle|^2]}$, where $G_{\rm pop}^2(t,\tau) = \langle b^\dagger b(t)\rangle \langle b^\dagger b(t+\tau)\rangle$, and $g^1(t,\tau)$ and $g^2(t,\tau)$ are the first- and second-order correlation functions, modeled as $g^1(t,\tau) = \langle b^\dagger(t) b(t+\tau)\rangle$ and $g^2(t,\tau) = \langle b^\dagger(t) b^\dagger(t+\tau) b(t+\tau) b(t)\rangle$. The brightness, defined as the emitted cavity photon number per pulse, is $\beta = \kappa \int_0^T \langle b^\dagger b(t)\rangle dt$. The probability of two or more photon emission for a single pulse excitation is given by $\tilde{P}_{2} = \frac{\int_0^T dt \int_0^T d\tau , g^2(t,\tau)}{\int_0^T dt \int_0^T d\tau , \langle b^\dagger(t)b(t)\rangle \langle b^\dagger(t+\tau)b(t+\tau)\rangle}$~\cite{ref55}. The upper limit of integration $T$ is assumed to be sufficiently long for the quantum dot to relax to its ground state. Moreover, as explained earlier, we assume that the probability of three or more photons is zero, i.e., $\tilde{P}_{\ge 3} = 0$. The zero-, single- and two-photon emission probabilities are then $p_0 = 1-p_1 - p_2-p_{\ge 3}$, $p_1 = \beta - 2 \tilde{P}_2$, and $p_2 = \tilde{P}_2$, respectively.
\section{ APPENDIX B: Secure Key Rate for BB84 and Twin-Field QKD Protocols}{\label{B}}

The lower bound on the secure key rate for both QDS and PDS single-photon sources is given by
$R \ge q[Q_1(1 - H_2(e_1)) - f Q_\mu H_2(E_\mu)]$,
where $q$ denotes the basis reconciliation factor ($q = 1/2$ for symmetric basis choice and $q \approx 1$ for efficient BB84 protocols~\cite{ref56}), $Q_\mu = \sum_{n=0}^{\infty} Y_n P_\mu(n)$ is the overall gain, $E_\mu$ is the overall QBER, $e_1$ is the single-photon QBER, $f$ is the error-correction inefficiency, $H_2(\cdot)$ is the binary entropy function, and $Y_n$ is the conditional detection probability (yield) for an $n$-photon state~\cite{ref18}. The yield of an $n$-photon pulse is modeled as
$Y_n = Y_0 + (1-Y_0) [1 - (1 - \eta_d \eta_t)^n]$,
where $Y_0$ is the dark count probability, $\eta_d$ is the detector efficiency, and $\eta_t = 10^{- \alpha L / 10}$ is the channel transmission, with $\alpha$ the fiber loss coefficient and $L$ the channel length in km.

Moreover, $P_\mu(n) = \frac{e^{-\mu} \mu^n}{n!}$ represents the photon number distribution for Poisson-distributed sources such as weak coherent pulses, where $\mu$ is the average photon number. It is worth noting that down-conversion sources intrinsically exhibit thermal statistics~\cite{ref57}; however, due to limited timing resolution or multimode averaging, the observed statistics may appear Poissonian ~\cite{ref58}. The error rate for an $n$-photon state is
$e_n = \frac{e_0 Y_0 + e_d [1 - (1 - \eta_d \eta_t)^n]}{Y_n}$,
and the gain for each photon number is $Q_n = Y_n P_\mu(n)$. The overall error rate therefore becomes
$E_\mu = \frac{1}{Q_\mu} \sum_{n=0}^{\infty} e_n Q_n$.

\subsection{BB84 QKD Without Decoy States} {\label{B1}}
In the GLLP model for non-decoy BB84, all errors and losses are pessimistically attributed to single-photon components~\cite{ref18}. For $n \ge 2$, one assumes $Y_n = 1$ and $e_n = 0$. Therefore, the single-photon gain and error rate satisfy
$Q_1 \ge Q_\mu - \sum_{n \ge 2} P_\mu(n)$ and $e_1 \le \frac{E_\mu Q_\mu}{Q_1}$.
Since multi-photon events are assumed to be fully compromised, the secure communication distance is strongly limited.
\subsection{BB84 QKD With Decoy States} {\label{B2}}
We numerically simulated the BB84 QKD protocol in the limit of infinite decoy states, which overcomes the limitation of accurately estimating the single-photon yield and error rate by allowing Alice to vary the mean photon number $\mu$ of her transmitted pulses over a continuum of values~\cite{ref21}. Since an eavesdropper cannot distinguish signal from decoy pulses, the channel parameters $Y_1$ and $e_1$ can be tightly estimated. In the ideal limit of infinitely many decoy intensities, the single-photon gain satisfies $Q_1 \ge Y_1 P_\mu(1)$, and the single-photon error rate is accurately determined as $e_1 = \frac{e_0 Y_0 + e_d \eta_d \eta_t}{Y_1}$. This greatly enhances both the secret key rate and the maximum feasible communication distance compared to BB84 without decoy states.
\subsection{Twin-Field QKD With Decoy States} {\label{B3}}
In TF-QKD, the detection scheme is closely related to that of phase-encoded decoy-state QKD and consists of single-photon interference followed by threshold single-photon detection. This similarity allows the use of a secure key-rate formula analogous to that of decoy-state QKD, with additional parameters accounting for phase matching and the protocol duty cycle. Specifically, it was shown in a seminal work on TF-QKD that, provided the revelation of the global phase after Charlie’s measurement does not increase the eavesdropper’s information, the secure key rate of TF-QKD can be written as~\cite{ref34}: $R^{\rm TF} = \frac{d}{M}[Q_1^{\rm TF}(1 - H_2(e_1^{\rm TF})) - f Q_\mu^{\rm TF} H_2(E_Q^{\rm TF})]$, where the gain of the $n$-photon component is $Q_n^{\rm TF} = Y_n^{\rm TF} P_\mu(n)$, and the overall gain is $Q_\mu^{\rm TF} = \sum_{n=0}^{\infty} Y_n^{\rm TF} P_\mu(n)$. The corresponding error rates are $e_n^{\rm TF} = \frac{e_0 Y_0 + (e_d + e_s) [1 - (1 - \eta_d \sqrt{\eta_t})^n]}{Y_n^{\rm TF}}$, \quad
$E_Q^{\rm TF} = \frac{1}{Q_\mu^{\rm TF}} \sum_{n=0}^{\infty} e_n^{\rm TF} Q_n^{\rm TF}$, which have the same functional form as in the infinite-decoy-state QKD analysis. In TF-QKD, unlike BB84, each optical pulse propagates only half of the total distance between Alice and Bob; consequently, the channel transmittance $\eta_t$ is replaced by $\sqrt{\eta_t}$. Accordingly, the yield of an $n$-photon component is modeled as $Y_n^{\rm TF} = Y_0 + (1 - Y_0) [1 - (1 - \eta_d \sqrt{\eta_t})^n]$. The intrinsic error rate $e_s = \frac{1}{2} - \frac{\sin(2\pi/M)}{4\pi/M}$ arises from finite phase slicing, where $M$ denotes the number of phase slices used for phase matching and $d$ is the fraction of protocol rounds allocated to key generation. For finite-decoy-state TF-QKD, modified expressions for the gain and error rates must be employed~\cite{ref59}. To evaluate the secure key rate for a quantum-dot single-photon source (QDS) in both BB84 and TF-QKD, the Poisson photon-number distribution $P_\mu(n)$ is replaced by $P_\eta(n)$. The photon-number probabilities for a QDS are given by $P_\eta(0) = p_0 + p_1 (1 - \eta) + p_1 (1 - \eta)^2$, \quad
$P_\eta(1) = p_1 \eta + p_2 [1 - \eta^2 - (1 - \eta)^2]$, \quad
$P_\eta(n \ge 2) = 1 - P_\eta(0) - P_\eta(1)$~\cite{ref38}.
\vspace{0.3cm}
\section*{Acknowledgment}

P.K. acknowledges the grants (SRG/2023/000560) from ANRF (SERB) and NQM, DST, Government of India.

\subsection*{Disclosures}
The authors declare no conflicts of interest.

\subsection*{Data availability}
The data that support the findings of this study are available upon reasonable request from the authors.


\begin{thebibliography}{99}

\bibitem{ref1}
C.~H.~Bennett and G.~Brassard, “Quantum cryptography: public key distribution and coin tossing,” \href{https://doi.org/10.1016/j.tcs.2014.05.025}{Theor. Comput. Sci. \textbf{560}, 7–11 (2014)}.

\bibitem{ref2}
C.~H.~Bennett, F.~Bessette, G.~Brassard, L.~Salvail, and J.~A.~Smolin, “Experimental quantum cryptography,” \href{https://doi.org/10.1007/BF00191318}{J. Cryptol. \textbf{5}, 3–28 (1992)}.

\bibitem{ref3}
C.~H.~Bennett, “Quantum cryptography using any two nonorthogonal states,” \href{https://doi.org/10.1103/PhysRevLett.68.3121}{Phys. Rev. Lett. \textbf{68}, 3121–3124 (1992)}.

\bibitem{ref4}
C.~H.~Bennett, G.~Brassard, and N.~D.~Mermin, “Quantum cryptography without Bell’s theorem,” \href{https://doi.org/10.1103/PhysRevLett.68.557}{Phys. Rev. Lett. \textbf{68}, 557–559 (1992)}.

\bibitem{ref5}
A.~Acín, N.~Gisin, and L.~Masanes, “From Bell’s theorem to secure quantum key distribution,” \href{https://doi.org/10.1103/PhysRevLett.97.120405}{Phys. Rev. Lett. \textbf{97}, 120405 (2006)}.

\bibitem{ref6}
J.-G.~Ren, P.~Xu, H.-L.~Yong, et al., “Ground-to-satellite quantum teleportation,” \href{https://doi.org/10.1038/nature23675}{Nature \textbf{549}, 70–73 (2017)}.

\bibitem{ref7}
M.~Zahidy, M.~T.~Mikkelsen, R.~Müller, et al., “Quantum key distribution using deterministic single-photon sources over a field-installed fibre link,” \href{https://doi.org/10.1038/s41534-023-00800-x}{npj Quantum Inf. \textbf{10}, 2 (2024)}.

\bibitem{ref8}
H.-Z.~Chen, M.-H.~Li, Y.-Z.~Wang, et al., “Implementation of carrier-grade quantum communication networks over 10,000 km,” \href{https://doi.org/10.1038/s41534-025-01089-8}{npj Quantum Inf. \textbf{11}, 137 (2025)}.

\bibitem{ref9}
Y.-A.~Chen, Q.~Zhang, T.-Y.~Chen, et al., “An integrated space-to-ground quantum communication network over 4,600 kilometres,” \href{https://doi.org/10.1038/s41586-020-03093-8}{Nature \textbf{589}, 214–219 (2021)}.

\bibitem{ref10}
T.-Y.~Chen, X.~Jiang, S.-B.~Tang, et al., “Implementation of a 46-node quantum metropolitan area network,” \href{https://doi.org/10.1038/s41534-021-00474-3}{npj Quantum Inf. \textbf{7}, 134 (2021)}.

\bibitem{ref11}
B.~Baghdasaryan, F.~Steinlechner, and S.~Fritzsche, “Enhancing the purity of single photons in parametric down-conversion,” \href{https://doi.org/10.1103/PhysRevA.108.023718}{Phys. Rev. A \textbf{108}, 023718 (2023)}.

\bibitem{ref12}
X.~You and Y.-M.~He, “Developing quantum-dot single-photon sources with excellent performance by coupling asymmetric microcavities,” 
\href{https://doi.org/10.1103/PhysRevResearch.7.L012016}
{Phys. Rev. Res. \textbf{7}, L012016 (2025)}.

\bibitem{ref13}
P.~Kumar and A.~G.~Vedeshwar, “Phonon-assisted control of the single-photon spectral characteristics,” \href{https://doi.org/10.1103/PhysRevA.96.033808}{Phys. Rev. A \textbf{96}, 033808 (2017)}.

\bibitem{ref14}
S.~Park, K.M.~Azizur-Rahman, D.~Shima, et al., “Efficient single-photon emission via quantum-confined charge funneling to quantum dots,” 
\href{https://doi.org/10.1038/s43246-025-01017-5}
{Commun. Mater. \textbf{6}, 286 (2025)}.

\bibitem{ref15}
S.~Kikura, R.~Asaoka, M.~Koashi, and Y.~Tokunaga, “High-purity single-photon generation based on cavity QED,” 
\href{https://doi.org/10.1103/PhysRevResearch.7.013251}
{Phys. Rev. Res. \textbf{7}, 013251 (2025)}.

\bibitem{ref16}
H.-K.~Lo and H.~F.~Chau, “Unconditional security of quantum key distribution over arbitrarily long distances,” \href{https://doi.org/10.1126/science.283.5410.2050}{Science \textbf{283}, 2050–2056 (1999)}.

\bibitem{ref17}
D.~Gottesman and H.-K.~Lo, “Proof of security of quantum key distribution with two-way classical communications,” \href{https://doi.org/10.1109/TIT.2002.807289}
{IEEE Trans. Inf. Theory \textbf{49}, 457–475 (2003)}.

\bibitem{ref18}
D.~Gottesman, H.-K.~Lo, N.~Lütkenhaus, and J.~Preskill, “Security of quantum key distribution with imperfect devices,” \href{https://doi.org/10.1109/ISIT.2004.1365172}{Quantum Inf. Comput. \textbf{4}, 325–360 (2004)}.

\bibitem{ref19}
N.~Gisin, G.~Ribordy, W.~Tittel, and H.~Zbinden, “Quantum cryptography,” \href{https://doi.org/10.1103/RevModPhys.74.145}{Rev. Mod. Phys. \textbf{74}, 145–195 (2002)}.

\bibitem{ref20}
G.~Brassard, N.~Lütkenhaus, T.~Mor, and B.C.~Sanders, “Limitations on practical quantum cryptography,” \href{https://doi.org/10.1103/PhysRevLett.85.1330}{Phys. Rev. Lett. \textbf{85}, 1330–1333 (2000)}.

\bibitem{ref21}
H.-K.~Lo, X.~Ma, and K.~Chen, “Decoy state quantum key distribution,” \href{https://doi.org/10.1103/PhysRevLett.94.230504}{Phys. Rev. Lett. \textbf{94}, 230504 (2005)}.

\bibitem{ref22}
W.-Y.~Hwang, “Quantum key distribution with high loss: Toward global secure communication,” \href{https://doi.org/10.1103/PhysRevLett.91.057901}{Phys. Rev. Lett. \textbf{91}, 057901 (2003)}.

\bibitem{ref23}
V.~Scarani, H.~Bechmann-Pasquinucci, N.J.~Cerf, et al., “The security of practical quantum key distribution,” \href{https://doi.org/10.1103/RevModPhys.81.1301}{Rev. Mod. Phys. \textbf{81}, 1301–1350 (2009)}.

\bibitem{ref24}
F.~Xu, X.~Ma, Q.~Zhang, H.-K.~Lo, and J.-W.~Pan, “Secure quantum key distribution with realistic devices,” \href{https://doi.org/10.1103/RevModPhys.92.025002}{Rev. Mod. Phys. \textbf{92}, 025002 (2020)}.

\bibitem{ref25}
X.~Ma, B.~Qi, Y.~Zhao, and H.-K.~Lo, “Practical decoy‑state quantum key distribution,” \href{https://doi.org/10.1103/PhysRevA.72.012326}{Phys. Rev. A \textbf{72}, 012326 (2005)}.

\bibitem{ref26}
Y.~Zhang, X.~Ding, Y.~Li, L.~Zhang, \textit{et al.}, 
``Experimental single-photon quantum key distribution surpassing the fundamental weak coherent-state rate limit,'' 
\href{https://doi.org/10.1103/PhysRevLett.134.210801}
{Phys. Rev. Lett. \textbf{134}, 210801 (2025).}

\bibitem{ref27}
X.~Ding, Y.-P.~Guo, M.-C.~Xu, et al., 
"High-efficiency single-photon source above the loss-tolerant threshold for efficient linear optical quantum computing,"
\href{https://doi.org/10.1038/s41566-025-01639-8}
{Nat. Photon. \textbf{19}, 387 (2025).}

\bibitem{ref28}
H.~Wang, Y.-M.~He, T.-H.~Chung, \textit{et al.}, 
"Towards optimal single-photon sources from polarized microcavities," 
\href{https://doi.org/10.1038/s41566-019-0494-3}
{Nat. Photon. \textbf{13}, 770 (2019).}

\bibitem{ref29}
C.~L.~Morrison, M.~Rambach, Z.~X.~Koong, \textit{et al.}, 
"A bright source of telecom single photons based on quantum frequency conversion,"
\href{https://doi.org/10.1063/5.0045413}
{Appl. Phys. Lett. \textbf{118}, 174003 (2021).} 

\bibitem{ref30}
F.~B.~Barnes, R.~G.~Pousa, C.~L.~Morrison, \textit{et al.}, "Decoy-state quantum key distribution over 227 km with a frequency-converted telecom single-photon source," \href{https://arxiv.org/abs/2512.05101}{arXiv:2512.05101 (2025)}.

\bibitem{ref31}
Y.~Bloom, Y.~Ordan, T.~Levin, \textit{et al.},
"Decoy-State and Purification Protocols for Superior Quantum Key Distribution with Imperfect Quantum-Dot-Based Single-Photon Sources: Theory and Experiment,"
\href{https://doi.org/10.1103/7fdd-m92n}
{PRX Quantum \textbf{6}, 030332 (2025).}

\bibitem{ref32}
Z.~Tang, K.~Wei, O.~Bedroya, L.~Qian, and H.-K.~Lo,
"Experimental measurement-device-independent quantum key distribution with imperfect sources,"
\href{https://doi.org/10.1103/PhysRevA.93.042308}
{Phys. Rev. A \textbf{93}, 042308 (2016).}

\bibitem{ref33}
Y.-M.~Xie, Y.-S.~Lu, C.-X.~Weng \textit{et al.}, "Breaking the rate-loss bound of quantum key distribution with asynchronous two-photon interference," 
\href{https://doi.org/10.1103/PRXQuantum.3.020315}
{PRX Quantum \textbf{3}, 020315 (2022).}

\bibitem{ref34}
M.~Lucamarini, Z.~L.~Yuan, J.~F.~Dynes, and A.~J.~Shields, “Overcoming the rate–distance limit of quantum key distribution without quantum repeaters,” \href{https://doi.org/10.1038/s41586-018-0066-6}{Nature \textbf{557}, 400 (2018)}.

\bibitem{ref35}
X.~Ma, P.~Zeng, and H.~Zhou, “Phase-matching quantum key distribution,” \href{https://doi.org/10.1103/PhysRevX.8.031043}{Phys. Rev. X \textbf{8}, 031043 (2018)}.

\bibitem{ref36}
Z.-W.~Yu, X.-L.~Hu, C.~Jiang, et al., “Sending-or-not-sending twin-field quantum key distribution in practice,” \href{https://doi.org/10.1038/s41598-019-39225-y}{Sci. Rep. \textbf{9}, 3080 (2019)}.

\bibitem{ref37}
X.-L.~Hu, C.~Jiang, Z.-W.~Yu, and X.-B.~Wang, “Sending‑or‑not‑sending twin‑field protocol for quantum key distribution with asymmetric source parameters,” \href{https://doi.org/10.1103/PhysRevA.100.062337}{Phys. Rev. A \textbf{100}, 062337 (2019)}.

\bibitem{ref38}
H.-L.~Yin and Z.-B.~Chen, “Coherent‑state‑based twin‑field quantum key distribution,” \href{https://doi.org/10.1038/s41598-019-50429-0}{Sci. Rep. \textbf{9}, 14918 (2019)}.

\bibitem{ref39}
M.~Bozzio, M.~Vyvlečka, M.~Cosacchi, A.~Rastelli, and V.~Zwiller, “Enhancing quantum cryptography with quantum dot single‑photon sources,” \href{https://doi.org/10.1038/s41534-022-00626-z}{npj Quantum Inf. \textbf{8}, 104 (2022)}.

\bibitem{ref40}
P.~Laccotripes, T.~Müller, R.~M.~Stevenson, et al., “Spin–photon entanglement with direct photon emission in the telecom C band,” \href{https://doi.org/10.1038/s41467-024-53964-1}{Nat. Commun. \textbf{15}, 9740 (2024)}.

\bibitem{ref41}
M.~Wasiluk, H.~Janowska, A.~Musiał, et al., “Probing electron spin dynamics in single telecom InAs(P)/InP quantum dots using the Hanle effect,” 
\href{https://doi.org/10.1063/5.0291044}
{Appl. Phys. Lett. \textbf{127}, 204001 (2025)}.

\bibitem{ref42}
X.-B.~Liu, S.-T.~Lyu, and H.-L.~Yin,
“Quantum dot source-based twin-field quantum key distribution,”
\href{https://doi.org/10.1364/OL.579134}{Opt. Lett. \textbf{51}, 644 (2026)}.

\bibitem{ref43}
S.-T.~Lyu, R.-Z.~Liu, and H.-L.~Yin,
“Twin-field quantum key distribution with coherent single-photon state,”
\href{https://doi.org/10.15302/frontphys.2026.083204}{Front. Phys. \textbf{21}, 083204 (2026)}.

\bibitem{ref44}
J.~C.~Loredo, C.~Antón, B.~Reznychenko, et al.,
“Generation of non-classical light in a photon-number superposition,”
\href{https://doi.org/10.1038/s41566-019-0506-3}{Nat. Photonics \textbf{13}, 803 (2019)}.

\bibitem{ref45}
X.~You and Y.-M.~He, "Developing the quantum-dot single-photon sources with excellent performance by coupling asymmetric microcavity,"  \href{https://doi.org/10.1103/PhysRevResearch.7.L012016}
{Phys. Rev. Research \textbf{7}, L012016 (2025)}.


\bibitem{ref46}
K.~H.~Madsen, S.~Ates, J.~Liu, A.~Javadi, S.~M.~Albrecht, I.~Yeo, S.~Stobbe, and P.~Lodahl, "Efficient out-coupling of high-purity single photons from a coherent quantum dot in a photonic-crystal cavity," 
\href{https://doi.org/10.1103/PhysRevB.90.155303}
{Phys. Rev. B \textbf{90}, 155303 (2014)}.

\bibitem{ref47}
J.~D.~Mar, X.~L.~Xu, J.~J.~Baumberg, et al., “Bias-controlled single-electron charging of a self-assembled quantum dot in a two-dimensional-electron-gas-based $n$--$i$ Schottky diode,” \href{https://doi.org/10.1103/PhysRevB.83.075306}{Phys. Rev. B \textbf{83}, 075306 (2011)}.

\bibitem{ref48}
U.~M.~Gür, M.~Mattes, S.~Arslanagić, and N.~Gregersen, “Elliptical micropillar cavity design for highly efficient polarized emission of single photons,” \href{https://doi.org/10.1063/5.0041565}{Appl. Phys. Lett. \textbf{118}, 061101 (2021)}.

\bibitem{ref49}
A.~Majumdar, P.~Kaer, M.~Bajcsy, et al., “Proposed coupling of an electron spin in a semiconductor quantum dot to a nanoscale optical cavity,” \href{https://doi.org/10.1103/PhysRevLett.111.027402}{Phys. Rev. Lett. \textbf{111}, 027402 (2013)}.

\bibitem{ref50}
G.~Gillard, I.~M.~Griffiths, G.~Ragunathan, A.~Ulhaq, et al.,
“Fundamental limits of electron and nuclear spin qubit lifetimes in an isolated self-assembled quantum dot,”
\href{https://doi.org/10.1038/s41534-021-00378-2}{npj Quantum Inf. \textbf{7}, 43 (2021)}.

\bibitem{ref51}
C.~Emary, X.~Xu, D.~G.~Steel, S.~Saikin, and L.~J.~Sham,
“Fast initialization of the spin state of an electron in a quantum dot in the Voigt configuration,”
\href{https://doi.org/10.1103/PhysRevLett.98.047401}{Phys. Rev. Lett. \textbf{98}, 047401 (2007)}.

\bibitem{ref52}
P.~Kumar and T.~Nakajima,
“Fast and high-fidelity optical initialization of spin state of an electron in a semiconductor quantum dot using light-hole-trion states,”
\href{https://doi.org/10.1016/j.optcom.2016.03.006}{Opt. Commun. \textbf{370}, 103 (2016)}.

\bibitem{ref53}
K.~A.~Fischer, K.~Müller, K.~G.~Lagoudakis, and J.~Vučković, “Dynamical modeling of pulsed two‑photon interference,” \href{https://doi.org/10.1088/1367-2630/18/11/113053}{New J. Phys. \textbf{18}, 113053 (2016)}.

\bibitem{ref54}
L.~Hanschke, K.~A.~Fischer, S.~Appel, et al., “Quantum dot single‑photon sources with ultra‑low multiphoton probability,” \href{https://doi.org/10.1038/s41534-018-0092-0}{npj Quantum Inf. \textbf{4}, 43 (2018)}.

\bibitem{ref55}
C.~Gustin and S.~Hughes, “Pulsed excitation dynamics in quantum‑dot–cavity systems: Limits to optimizing the fidelity of on‑demand single‑photon sources,” \href{https://doi.org/10.1103/PhysRevB.98.045309}{Phys. Rev. B \textbf{98}, 045309 (2018)}.

\bibitem{ref56}
H.-K.~Lo, H.~F.~Chau, and M.~Ardehali, “Efficient quantum key distribution scheme and a proof of its unconditional security,” \href{https://doi.org/10.1007/s00145-004-0142-y}{J. Cryptol. \textbf{18}, 133 (2005)}.


\bibitem{ref57}
B.~Blauensteiner, I.~Herbauts, S.~Bettelli, et al., “Photon bunching in parametric down‑conversion with continuous‑wave excitation,” \href{https://doi.org/10.1103/PhysRevA.79.063846}{Phys. Rev. A \textbf{79}, 063846 (2009)}.


\bibitem{ref58}
H.~Takesue and K.~Shimizu, “Effects of multiple pairs on visibility measurements of entangled photons generated by spontaneous parametric processes,” \href{https://doi.org/10.1016/j.optcom.2009.10.008}{Opt. Commun. \textbf{283}, 276 (2010)}.

\bibitem{ref59}
Q.~Peng, J.-P.~Chen, T.~Xing, et al., “Practical security of twin‑field quantum key distribution with optical phase‑locked loop under wavelength‑switching attack,” \href{https://doi.org/10.1038/s41534-025-00963-9}{npj Quantum Inf. \textbf{11}, 7 (2025)}.

\end{thebibliography}
\end{document}